\documentclass[pre,amsmath,amssymb,twocolumn,superscriptaddress,showpacs]{revtex4-1}

\usepackage{amsmath,amssymb}
\usepackage[usenames]{color}
\usepackage{amssymb}
\usepackage{grffile}
\usepackage[pdftex]{graphicx}
\usepackage{amsmath, amstext, amssymb, amsfonts, amsxtra}
\usepackage{textcomp}
\usepackage{xspace}
\usepackage[colorlinks]{hyperref}

\def \tr{{\mbox{tr~}}}

\def \l{{\lambda}}

\def \w{{\omega}}

\def \ell{{d}}
\newcommand{\dop}{\hat{d}}
\newcommand{\ddop}{\hat{d}^{\dagger}} 
\newcommand{\xop}{\hat{X}}
\newcommand{\Dop}{\mathcal{D}}

\newcommand{\ran}{\rangle}

\newcommand{\im}{{\rm i}} 
\newcommand{\rhop}{\hat{\rho}}

\newcommand{\Hopi}{\hat{H}_{\rm int}}
\newcommand{\Hz}{\hat{H}_0}
\newcommand{\sop}{\hat{\sigma}} 
\newcommand{\sopp}{\hat{\sigma}^+} 
\newcommand{\sopm}{\hat{\sigma}^-}

\newcommand{\kb}[2]{| #1 \rangle\langle #2 | }
\newcommand{\kbj}[2]{|\! #1 \rangle_j\langle #2 \!| }
\newcommand{\proj}[1]{| #1 \rangle\langle #1 | }
\newcommand{\projj}[1]{| #1 \rangle_j\langle #1 | }
\newcommand{\ave}[1]{\langle #1 \rangle }
\newcommand{\pare}[1]{\left( #1 \right) }
\newcommand{\spare}[1]{\left[ #1 \right] }
\newcommand{\half}[1]{\frac{ #1 }{2} }
  
\newcommand{\Del}{\nu} 
\newcommand{\hc}{{\rm H.c.}}
\newcommand{\id}{\hat{\mathbb{I}}} 
\newcommand{\lt}{\lambda}
\newcommand{\Lz}{\mathcal{L}_0}
\newcommand{\Vp}{\mathcal{V}} 
\newcommand{\Pro}{\mathcal{P}} 
\newcommand{\Leff}{\mathcal{L}_{\rm eff}}
 
\newcommand{\Ss}{\hat{\mathcal{S}}}
\newcommand{\upprojj}{| \! \uparrow \rangle_j\langle \uparrow \! | }
\newcommand{\downprojj}{| \! \downarrow \rangle_j\langle \downarrow \! | }

\newcommand{\sutd}{Singapore University of Technology and Design, 8 Somapah Road, 487372 Singapore} 
\newcommand{\majulab}{MajuLab, CNRS-UNS-NUS-NTU International Joint Research Unit, UMI 3654, Singapore}

\begin{document}

\title{Converting heat into directed transport on a tilted lattice}       
%\title{Classical and quantum aspects of a minimal self-contained heat engine}       
%\title{Self-contained energy trasducer coupled to a particle on a linear slope}       
\author{Colin Teo} 
\affiliation{\sutd} 
\author{Ulf Bissbort} 
\affiliation{\sutd} 
\author{Dario Poletti}
\affiliation{\sutd} 
\affiliation{\majulab}

\begin{abstract}
We present a self-contained engine, which is made of one or more two-level systems, each of which is coupled to a single bath, as well as to a common load composed of a particle on a tilted lattice. We show that the energy and the entropy absorbed by the spins are transferred to the particle thus setting it into upward motion at an average constant speed, even when driven by a single spin connected to a single bath. When considering an ensemble of different spins, the velocity of the particle is larger when the tilt is on resonance with any of the spins' energy splitting. Interestingly, we find regimes where the spins' polarization enters periodic cycles with the oscillation period being determined by the tilt of the lattice.   
\end{abstract}

\maketitle

%%%%%%%%%%%%%%%%%%%%%%%%%%%%%%%%%%%%%%%%%%%%
%%%%%%%%%%%%%%%%%%%%%%%%%%%%%%%%%%%%%%%%%%%%
% general introduction 

%Beauty and importance of a simple model which already shows such an interesting physics.   

%%%%%%%%%%%%%%%%%%%%%%%%%%%%%%%%%%%%%%%%%%%%
%%%%%%%%%%%%%%%%%%%%%%%%%%%%%%%%%%%%%%%%%%%%
% general introduction 

%Connect to atom-tronics?  
Energy transport and conversion at the nanoscale are areas of increasing interest due to possible applications in future technologies (for reviews see, e.g.,~\cite{GiazottoPekola2006, Shakouri2011,Dubi2011,Sothmann2014,Benenti2016, MuhonenPekola2012, Seifert2012, Kosloff2013, Gelbwaser2015, Vinjanampathy2015,Benenti2016b}). 
To study the energy conversion from heat (originating from a bath) to useful work, three types of quantum heat engines are usually considered: 1) continuously driven, 2) self-contained and 3) stroke engines. 
In continuously driven engines, the working fluid is constantly in contact with the baths and a periodic driving \cite{Gelbwaser2015}. In stroke engines the coupling of the system to the different baths alternates in a periodic fashion, and, for certain strokes, the engine may not be coupled to any of the baths \cite{Seifert2012, Kosloff2013, Vinjanampathy2015}. 
In both cases, the ``load'' to which the engine is coupled, is modeled by a periodic modulation of the Hamiltonian of the system, implying the absence of entropy flow between the engine and the load. Hence the energy transfer is exclusively work. 

Self-contained engines, on the other hand, include a quantum-mechanical description of the load and the equations of motion are time-independent. The first example of this type of engine was given in \cite{ScovilSchulzDuBois1959} and important advances were recently made in \cite{Youssef2009, Linden2010, Skrzypczyk2011}, where minimal examples of self-contained engines and refrigerators were proposed. Such minimal models give a deep insight into the fundamental mechanisms for the emergence of physical phenomena and are thus of great importance. Moreover, such self-contained systems allow the study of correlations which naturally arise between the working fluid and the load, since \emph{both} energy and entropy are transferred from engine to load in general. 

\begin{figure}
\includegraphics[width=0.9\columnwidth]{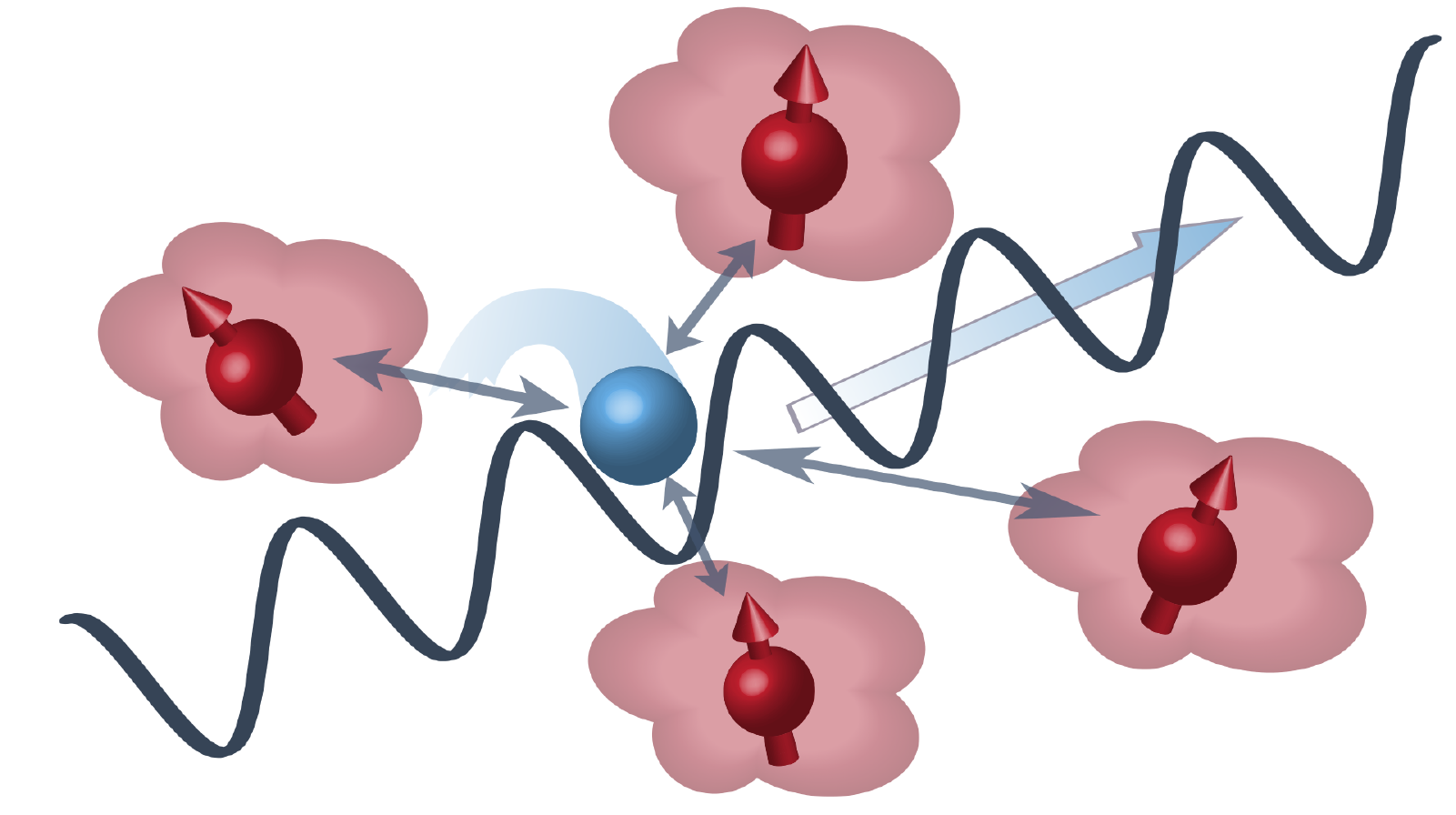}
\caption{(Color online) Pictorial representation of the system studied. An ensemble of spins each coupled to a single bath (depicted by clouds around the spins), and to a particle on a tilted lattice.} \label{fig:fig1} 
\end{figure}

Here we study a system composed of one or more spin $1/2$ (the engine) and a particle on an infinite tilted lattice (the load) as shown in Fig.~\ref{fig:fig1}. Each individual spin is coupled to both a single bath and the particle, allowing for a net transfer of energy from the baths to the particle. We will first focus on the behavior of the load and then on that of the spins. We show that, because of the energy transfer through the spins, the particle can be set to move up the tilted lattice corresponding also to an energy increase of the particle. We use two effective analytical approaches to understand the dynamical behavior: first, we use an ansatz for which the spins are uncorrelated from the load. Second, we account for small amounts of coherence and correlations on very short time scales. We also perform exact, numerical simulations which go beyond the regime of validity of the analytical approaches. 
By studying the transport of the particle via its velocity (both for the propagation and its spread) and its entropy change, we show that at sufficiently long times, the motion of the particle is governed by a biased classical diffusive dynamics. In particular, we uncover an upper bound for the particle velocity, which only depends on the coupling to the bath. As for the spins dynamics, we show that it can converge to a limit cycle when the particle has some initial coherence.

%%%%%%%% Figure1   

{\it Model}: The dynamics of the overall system, described by a density operator $\rhop$, is given by a time-independent master equation in Lindblad form \cite{GoriniSudarshan1976,Lindblad1976} 
\begin{align}      
\frac{d\rhop}{dt}=\Lz(\rhop)+\Vp(\rhop) \label{eq:masterequation},
\end{align}       
where $\Lz = -(\im/\hbar) \left[\Hz,\rhop\right]+\Dop(\rhop)$ describes the dynamics of the spins and the particle independently. Here 
\begin{align}
\Hz & = \hbar\Del \xop + \sum_j\frac{\hbar\w_j}{2} \sop_{z,j}  \label{eq:H0}     
\end{align}
and each spin could be coupled to different baths, described by a dissipator in Lindblad form    
\begin{align}
\Dop(\rhop) =& \sum_j\lambda^-_j \left( \sopm_j \rhop \sopp_j - \frac{1}{2} \{ \sopp_j \sopm_j,\rhop\} \right) \nonumber \\
& +\lambda^+_j \left(\sopp_j \rhop \sopm_j - \frac{1}{2} \{ \sopm_j \sopp_j,\rhop\}\right).  \label{eq:lindblad}
\end{align} 
In Eq.~(\ref{eq:H0}) $\hbar\w_j$ is the energy difference between the two levels of the $j$-th spin, $\hbar\Del$ is the particle's energy difference between two levels of the tilted lattice, $\xop = \sum_n n \proj n$ and $\sop_{z,j}$ is the usual Pauli spin operator. The $\lambda^+_j$ and $\lambda^-_j$ in Eq.~(\ref{eq:lindblad}) are, respectively, raising and lowering rates of the $j$-th spin, while $\sopp_j$ and $\sopm_j$ are spin raising and lowering operators.
The interaction between the spins and the particle is described by a Jaynes-Cummings type coupling \cite{JaynesCummings1963}, $\Hopi=\sum_j \hbar g_j\left(\dop\sopp_j + \ddop \sopm_j\right)$, giving
\begin{align}
\Vp(\rhop) & = -\frac{\im}{\hbar} \left[\Hopi,\rhop\right]. \label{eq:JCHam}     
\end{align} 
in Eq.(\ref{eq:masterequation}). Here, $\dop = \sum_n \kb{n}{n+1}$, and $\hbar g_j$ is the strength of the coupling between the $j$-th spin and the particle.

We note that in the system described by Eqs.(\ref{eq:masterequation}-\ref{eq:JCHam}) $\spare{\dop,\ddop} = 0$ and $\spare{\dop,\xop}=\dop$, which results in    
\begin{align}
\ave{\dop(t)} &= d_0 e^{-\im \nu t}\label{eq:bt},
\end{align}
where we used $\ave{\dop(t = 0)} = d_0$. 
The velocity of the particle, given by the time-derivative of the position, can be obtained from the Heisenberg equations of motion for $\xop$ and is given by 
\begin{align}
\frac{d\ave{\xop}}{dt} = -\im \sum_j g_j \ave{\sopp_j \dop\,} + \hc \label{eq:dxopdt}
\end{align}

{\it Separable ansatz}: It is insightful to study the dynamics of this system using a separable ansatz for which the total density operator is constrained to be a direct product of the spin and the load 
\begin{equation}
\rhop(t) \approx  \rhop_l(t) \otimes_j \rhop_{s,j}(t), \label{eq:sep}    
\end{equation}
throughout the evolution. This constraint allows no correlations to exist between the spin and the load. One can obtain the equations of motion, which retain the separable form of the density matrix in Eq.~(\ref{eq:sep}) by decoupling $\dop\sopp_j \approx \langle \dop \rangle \sopp_j + \dop \langle \sopp_j\rangle - \langle \dop\rangle \langle \sopp_j\rangle$ in $\Hopi$ (the conjugate term is decoupled analogously). This leads to the coupled set of equations
\begin{align}
\dot \rhop_{s,j}(t) &= -\im \left[\frac{\w_j}{2} \sop_{z,j} + g_j \left(\sopp_j d_0 e^{-\im \nu t}+\sopm_j d_0^* e^{\im \nu t}\right) \! , \; \rhop_{s,j} \right] \nonumber\\ & + \Dop\left(\rhop_{s,j}\right) \label{eq:ucs}\\
\dot \rhop_l(t) &= -\im \sum_j \left[\w_l \xop +g_j \left(\ave{\sopp_j} \dop +\ave{\sopm_j}\ddop\right) \! , \;\rhop_l\right]. \label{eq:ucl}
\end{align}
Note, that we have used Eq.~(\ref{eq:bt}) in Eq.~(\ref{eq:ucs}), which allows us to completely solve for the spin dynamics (see \cite{AppendixA} for details).  
Using the solution for the spin dynamics, we obtain the analytic expression for asymptotic value of the particle velocity within this separable ansatz approach      
\begin{equation}
\lim_{t\to \infty} \left. \frac{d\ave{ \xop}}{dt}\right|_{\rm sep} = \sum_j  \frac{ |d_0|^2 g_j^2\left(\lambda^+_j- \lambda^-_j\right)}{2 |d_0|^2 g_j^2 + \lt^2_j + \pare{\w_j - \Delta}^2}, \label{eq:velsep}
\end{equation}
where we used $\lt_j=(\lambda^+_j +\lambda^-_j)/2$. 
It should be noted that Eq.~(\ref{eq:velsep}), derived without including any correlation between the spin and the load, predicts that the load can move at constant velocity provided at time $t=0$ there is coherence in the load (i.e. $\ave{\dop(0)}\ne 0$). 

{\it Adiabatic elimination}: For small $g_j$ the evolution due to the interaction between the spins and the load, described by Eq.~(\ref{eq:JCHam}), can be analyzed perturbatively with an effective Linbladian super-operator $\Leff=-\Pro_0\Vp \Lz \Vp\Pro_0$ (see \cite{CiracPhillips1992, GarciaRipollCirac2009, PolettiKollath2012, PolettiKollath2013, SciollaKollath2015}). Here, $\Pro_0$ projects the density operator onto the null space of $\Lz$, which is given by any density operator of the form  
\begin{align}
\rhop = \pare{\sum_n P_n \proj{n} }\bigotimes_j \pare{p^+_j \upprojj + p^-_j \downprojj}, \label{eq:null}
\end{align}
where $p^a_j=\l^a_j/(2\l_j)$ and $a=\pm$ denotes the probability for the spin to be in either the up or down state in the absence of coupling, $\projj{b}$ projects on the $b$ state of the $j$-th spin, and $P_n$ is the probability that the particle occupies site $n$. Applying the effective Linbladian $\Leff$ to any density operator described by Eq.~(\ref{eq:null}) and subsequently tracing out the spins' degrees of freedom, one obtains a diffusion equation for $P_n$ (see \cite{appendixB} for more details) 
\begin{align}
\dot{P}_n=\sum_j g_j^2\frac{ \spare{ \lambda^+_j P_{n-1} - 2\lt_j P_n + \lambda^-_j P_{n+1}  }}{\lt_j^2+\pare{\Del - \w_j}^2}. \label{eq:padi}   
\end{align} 
It directly follows that this biased classical diffusion equation predicts a linear increase in time of the particle position $\ave{\xop}_{\rm ae}$ with the velocity
\begin{align}
\left.\frac{d\ave{\xop}}{dt}\right|_{\rm ae}=\sum_j g_j^2\frac{\lambda^+_j-\lambda^-_j}{\lt_j^2+\pare{\Del - \w_j}^2} \label{eq:xadi},
\end{align}
as well as a linear increase of the variance of the particle position $\Delta X^2_{\rm ae}=\ave{\xop^2}_{\rm ae}-\ave{\xop}^2_{\rm ae}= 2g^2\lt t/(\lt_j^2+(\Del-\w_j)^2)$. 
Unlike the previous scenario within the separable ansatz, the particle in this case also moves if the initial condition has no coherence $\ave{\dop(0)}=0$. Hence, the presence of coherence and correlations between the spin and the load, even when considered only on the perturbative level, leads to qualitatively different dynamical behavior.

{\it Velocity, variance, entropy and response to the load}: To go beyond the limitations of the analytical approaches discussed above, we now study the system numerically. In our simulations, we solved Eqs.~(\ref{eq:masterequation}-\ref{eq:JCHam}) using a sufficiently large lattice, such that the system's behavior is unaffected by the boundaries \cite{lowerbound}. 
We have used various numbers of spins and types of initial conditions, including scenarios in which there is a single spin which is entangled with the particle \cite{initialcondition}.

\begin{figure}
\includegraphics[width=\columnwidth]{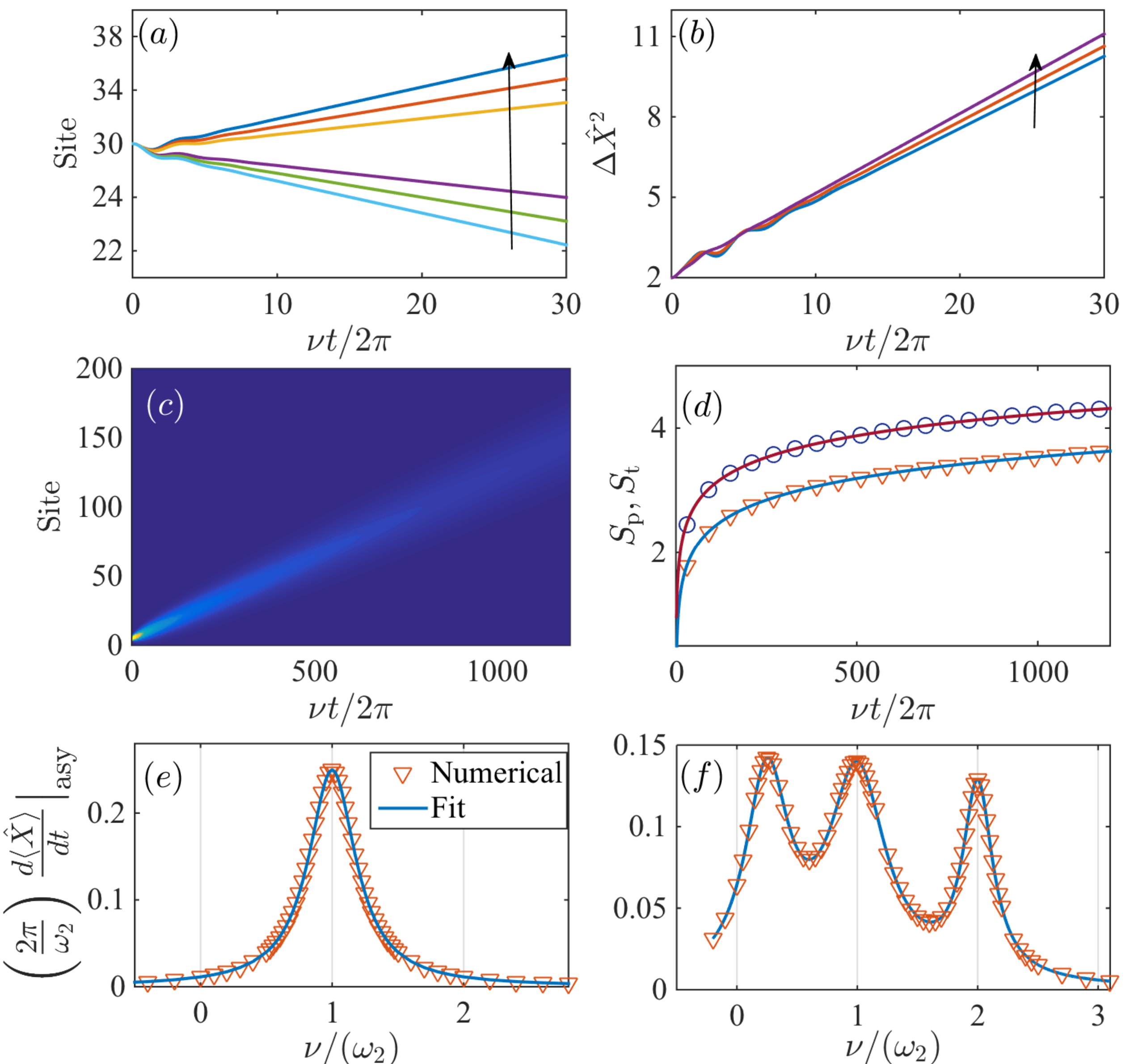}     
\caption{(Color online) (a) Average and (b) variance of the position of the particle vs time for different values of $\lambda^+-\lambda^-$. In (a) and (b) we only consider a single spin and keep $(\lambda^+_1+\lambda^-_1)/2=\l_1$ constant. 
The parameters used for (a) are, $\lambda^+_1/\nu = \{0.01,0.02,0.03,0.07,0.08,0.09\}$ while for (b), $\lambda^+_1/\nu = \{0.045,0.07,0.09\}$. In both (a,b) the arrow indicates the direction of increasing $\lambda^+_1$. (c) Density plot of $P_n$ versus position and time. (d) Total entropy $S_t$ (circles) and particle entropy $S_p$ (triangles) vs time. The continuous lines are fit with the function $y(t)=a+[\ln(t)]/2$ predicted by classical normal diffusion. In (a)-(d), we used $\lambda_j = 0.05 \nu$, $\w_1 = \nu$ and $g_1 = 0.15 \nu$. In (c) and (d) $\lambda^+ = 0.07 \nu$. In (e) and (f) we show the particle velocity vs $\w-\Del$. (e) shows two identical spins coupled by $g_1=g_2$ while (f) shows the results for three different spins. Plotted are both the asymptotic results of a full numerical simulation as well as fits using a sum of Lorentzian lineshapes, fitted using a least squares algorithm. 
The parameters in (e) are $\w_1=\w_2$ and $g_1=g_2=0.15\w_2$. In (f) instead $\w_1=\w_2/4$, $\w_3=2\w_2$, $g_1/\w_2=0.15$, $g_2/\w_2=0.2$ and $g_3/\w_2=0.1$. In both (e) and (f) $\l^+_j/\w_2=0.071$ and $\lt_j/\w_2=0.05$ for every $j$.} \label{fig:fig2} 
\end{figure}

As predicted by both the separable ansatz and the adiabatic elimination approach, the velocity of the particle asymptotically approaches a constant value, as highlighted in Fig.~\ref{fig:fig2}(a). As shown, the asymptotic velocity is reached after a short time ($\sim  20 \pi/\nu$), this was additionally verified with all the tested initial conditions. The variance $\Delta\xop^2=\ave{\xop^2}-\ave{\xop}^2$ versus time, shown in Fig.~\ref{fig:fig2}(b), also grows linearly as predicted by the adiabatic elimination approach. A typical density plot of the probability of finding the particle at position $n$ versus time is depicted in Fig.~\ref{fig:fig2}(c), where we have considered a single spin with coupling strength $g_1=0.15\nu$, which is beyond the regime of validity of the adiabatic elimination approach \cite{validity}. Interestingly, given the evolution of the average position and variance of the particle, the dynamics is still very well described, at least asymptotically, by biased classical diffusion.      

It is important to point out that the entropy of the overall system, and in particular that of the particle, increases continuously in time due to the transfer of entropy from the bath to the load. Fig.~\ref{fig:fig2}(d) shows both the entropy of the total system, $S_t=-\tr\spare{\rhop \ln\pare{\rhop}}$  (circles), and the entropy of the particle, $S_p=-\tr\spare{\rhop_p \ln\pare{\rhop_p}}$ (triangles), as functions of time, for the case of a single spin driving the particle. Here $\rhop_p=\tr_{\!\!s\!}\spare{\rhop}$ is the reduced density matrix of the particle alone, after having traced out the spin. The evolution of the entropy versus time can be predicted by computing the entropy of a continuous Gaussian distribution $p(x,t)=\pare{1/\sqrt{4\pi D t}}e^{-(x-v)^2/(4Dt)}$ obtained as a solution of a classical biased normal diffusion $\partial p /\partial t= D\partial^2 p /\partial x^2 + v \partial p /\partial x$, where $D$ is the diffusion constant and $v$ the velocity. One thus get  
\begin{align} 
S_c&=-\int_{-\infty}^{\infty} p(x,t) \ln\spare{p(x,t)} = \frac 1 2 \spare{1+\ln(4\pi Dt)} .    
\end{align}   
In Fig.~\ref{fig:fig2}(d) we show that a fit with a function $y=a+\ln(t)/2$ (continuous lines), where $a$ is the fitting parameter, agrees very well with both numerically evaluated entropies. This also shows that, at long times, the two entropies only differ by a constant term, i.e. all the entropy is transferred from the bath to the particle.  
     
The velocity of the particle is also dependent on the difference between the frequency of the spin $\w$ and that of the load $\Del$, and is largest at resonance. In particular, the velocity vs $\Del$ curve has a Lorentzian shape as predicted both by the separable ansatz and the adiabatic elimination (see Eqs.~(\ref{eq:velsep}) and (\ref{eq:xadi})). In Fig.~\ref{fig:fig2}(e), we show the behaviour for two identical spins with the same $\w_j$ and $g_j$. For this system only a single peak is present for $\Del=\w_2$. In Fig.~\ref{fig:fig2}(f) we present the velocity vs $\Del$ curve for a system consisting of three different spins with frequencies $\w_1=\w_2/2$ and $\w_3=2\w_2$ (while the individual coupling strengths, $g_j$, are also different). The velocity then shows three different peaks for $\Del$ equal to any of these frequencies.  We have also added fits with the sum of Lorentzian lineshapes in both Figs.~\ref{fig:fig2}(e) and \ref{fig:fig2}(f) and observe remarkable agreement between the fits and numerical simulation. We note that no side-peaks or harmonics are present \cite{shorttime}. 

%%%%%%%% Figure2    
\begin{figure}
\includegraphics[width=\columnwidth]{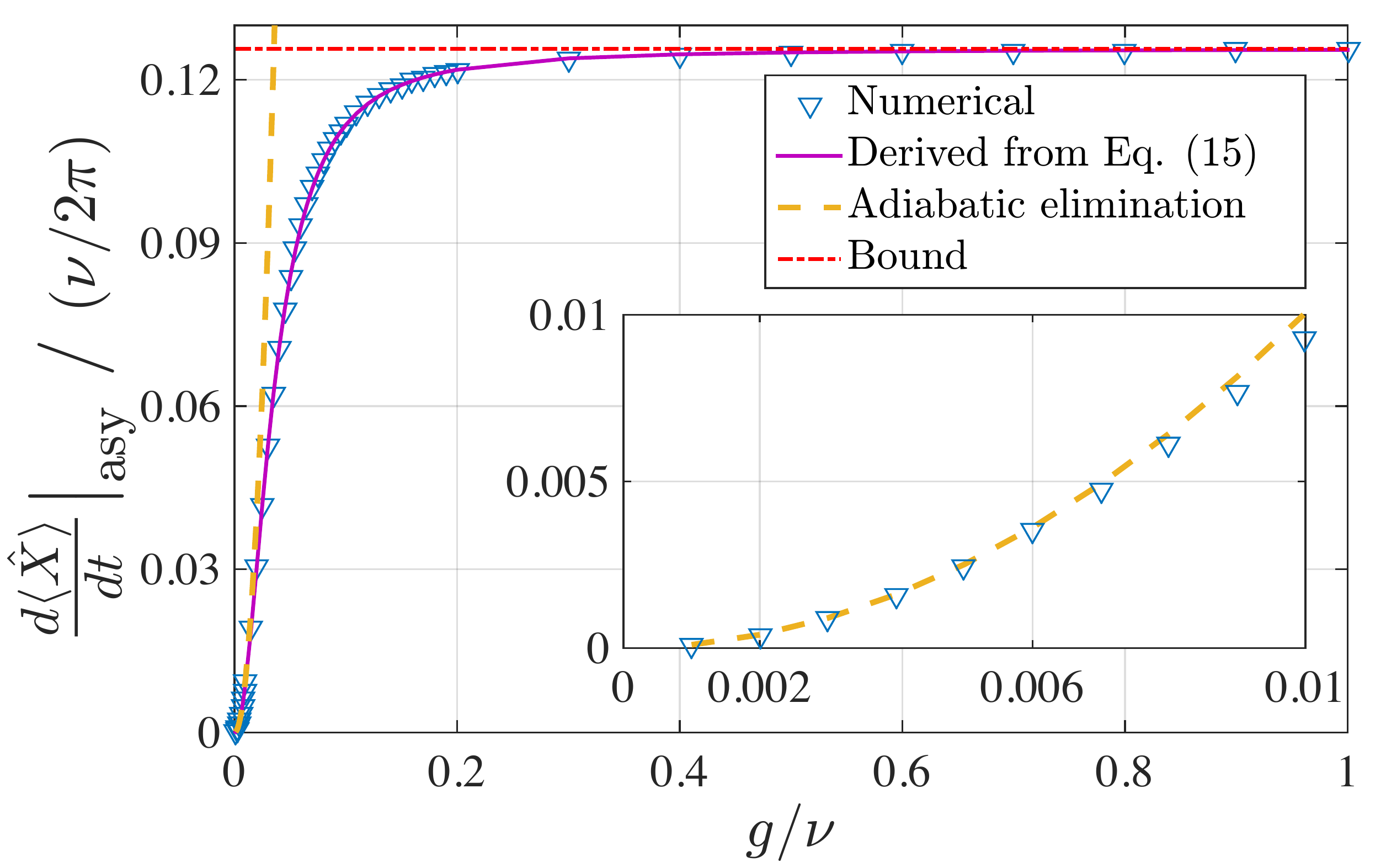}
\caption{(Color online) Asymptotic velocity of the particle vs coupling to the spin $g_1$. The filled circles (joined by a dashed line) represent the numerical results, the continuous line represent the asymptotic value $(\l^--\l^+)/2$ while the dashed line represent the prediction of the adiabatic elimination Eq.~(\ref{eq:xadi}). The inset shows a detail of Fig.~\ref{fig:fig3} which shows the accuracy of the prediction of Eq.~(\ref{eq:xadi}). In the plot, we have considered only a single spin with $\w_1 = \nu$, $\lambda_1^+ = 0.071\nu$, and $\lambda_1=0.05\nu$. 
} \label{fig:fig3} 
\end{figure}

The adiabatic elimination approach is qualitatively correct, however, as soon as the coupling between the spin and the load is not small compared to the dissipation with the baths, it becomes quantitatively inaccurate. In particular, both the velocity and the constant diffusion of the particle are over-estimated. In Fig.~\ref{fig:fig3}, we show the velocity of the particle vs the coupling strength $g_1$ (for this example we use a single spin). For small values of $g_1$, the velocity grows as $g_1^2$ and is exactly predicted by Eq.~(\ref{eq:xadi}), as shown by the dashed line. 
However, for larger values of $g_1$, the concavity of the curve changes and the velocity reaches an asymptotic value. To understand this, we can use that (i) the evolution of $\sop_{z,1}$, computed from Eqs.~(\ref{eq:masterequation}-\ref{eq:JCHam}), is given by $d\sop_{z,1}/dt=\im 2 g_1 \pare{\sop^+_1\dop - \sop^-_1\ddop} +\pare{\lambda^+_1-\lambda^-_1}\id -2 \lt_1\sop_{z,1}  $, (ii) that $d \ave{\xop}/dt$ follows Eq.~(\ref{eq:dxopdt}) and (iii) that $\ave{\sop}_{z,1}$ reaches an asymptotic value $\ave{\sop_{z,1}}_{\rm asy}$. We thus obtain             
\begin{align}
\left.\frac{d \ave{\xop}}{dt}\right|_{\rm asy}=\frac{\lambda^+_1-\lambda^-_1}{2} - \frac{\lt_1}{2}\ave{\sop_{z,1}}_{\rm asy}.  \label{eq:asymptotic}      
\end{align}
For $g_1=0$, we have $\ave{\sop_{z,1}}_{\rm asy}=\pare{\lambda^+_1-\lambda^-_1}/\lt_1 $, which implies that $d \ave{\xop}/dt|_{\rm asy}=0$. As $g_1$ increases, the asymptotic value of $\ave{\sop_{z,j}}$ approaches to zero and thus the particle will move at its maximum velocity $\pare{\lambda^+_1-\lambda^-_1}/2$ which is shown by the dot-dashed line in Fig.~\ref{fig:fig3}. We have compared the numerical results for the asymptotic velocity $d \ave{\xop}/dt|_{\rm asy}$ (triangles), with the prediction of Eq.~(\ref{eq:asymptotic}), where the respective $\ave{\sop_{z,j}}$ is computed by a numerical simulation of the full master equation, and we find excellent agreement (continuous line). 
%%%%%%%%% Figure3    
% 
%
%%%%%%%% Figure4               
\begin{figure}
\includegraphics[width=\columnwidth]{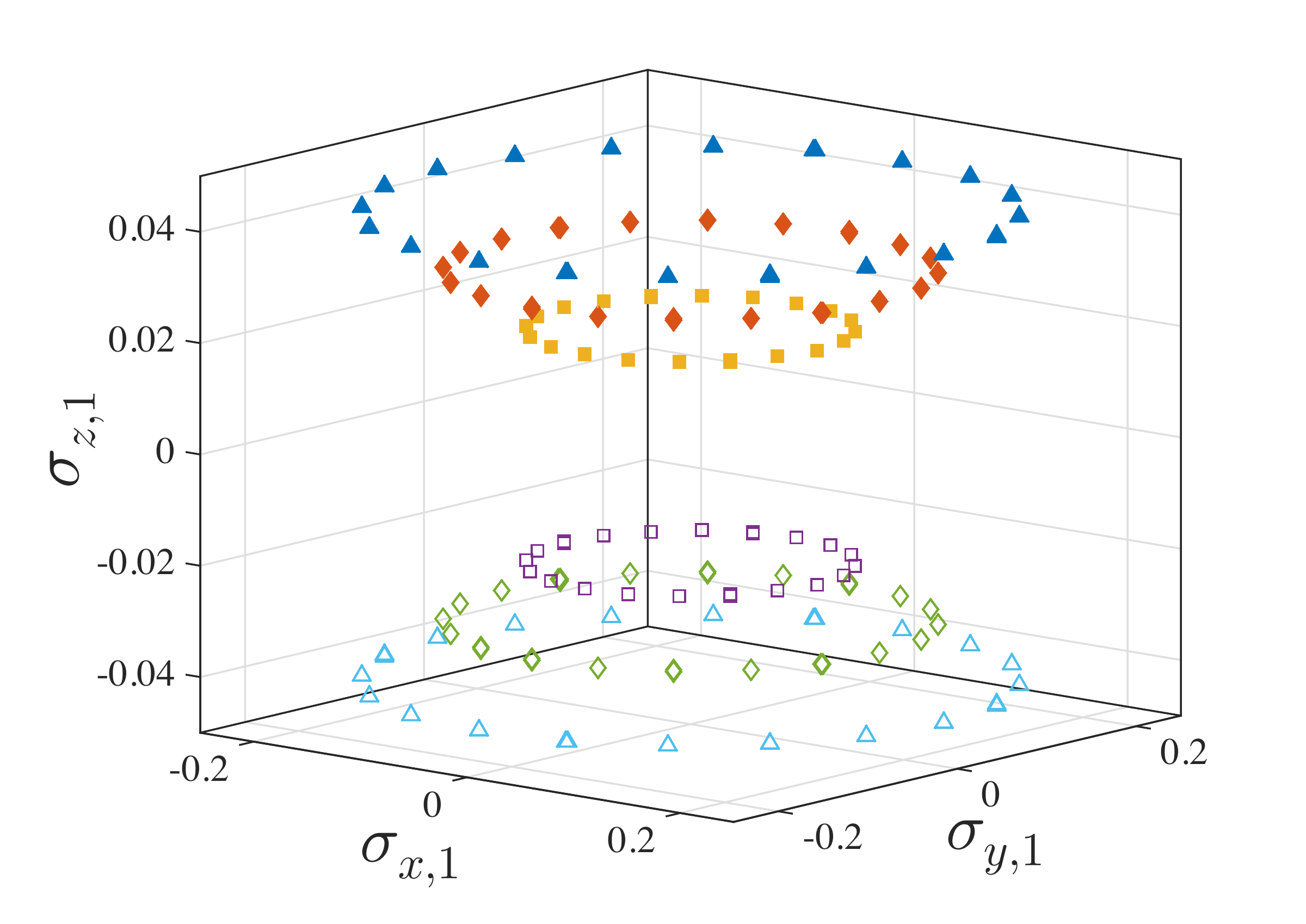}
\caption{(Color online) Asymptotic behavior of $\ave{\sop_{x,j}}$, $\ave{\sop_{y,j}}$ and $\ave{\sop_{z,j}}$ for different values of $\lambda^+_1 - \lambda^-_1$. From bottom to top we have used $\l^+_1/\nu=\{ 0.01, 0.02, 0.03, 0.07,0.08,0.09\}$ while keeping $\lt_1/\nu=	0.05$. The other parameters are $\w_1/\nu = 1$ and $g/\nu = 0.15$.} \label{fig:fig4} 
\end{figure}

{\it Limit cycle}: We now move our focus on the spin(s) part of the system (i.e. the engine) and study its evolution. Interestingly, a limit cycle behavior for the spins can emerge naturally. If the initial particle state has non-zero $d_0$, then the operators $\ave{\sop_{x,j}}$ and $\ave{\sop_{y,j}}$ exhibit periodic motion, while $\ave{\sop_{z,j}}$ reaches an asymptotic value. This is shown in Fig.~\ref{fig:fig4}, where we consider a system with a single spin and plot a number of discrete, equitemporally spaced points $[\ave{\sop_{x,1}}, \ave{\sop_{y,1}}, \ave{\sop_{z,1}}]$ lying on the continuous cyclic trajectories for different values of $\lambda^+_1$ and $\lambda^-_1$, while keeping $\lambda_1$ constant. These points are obtained after evolving the system for a sufficient time to reach the limit cycle (often a time of a dozen of $2\pi/\omega$ suffices). The period of the motion is given by $T_{\rm d}=2\pi/\Del$, and can be explained by an effective periodic driving due to the free evolution of $\ave{\hat d \,}$ as discussed in Eq.~(\ref{eq:bt}). Interestingly, this qualitatively coincides with the prediction of the separable ansatz. The amplitude of the oscillations decreases for decreasing coupling $g_1$ (and vanishes as $g_1\rightarrow 0$) while it increases for larger $|\lambda^+_1 - \lambda^-_1|$. 
In the presence of a limit cycle, the separable ansatz given by Eq.~(\ref{eq:sep}), gives an interesting insight. Since the spins are connected to a bath and to a periodic driving, it can be understood as a continuously driven engine. In particular, this set-up described within a separable ansatz is analogous to that described by Scully et al. \cite{ScullyWalther2003}, for which the presence of coherence allows the engine to work even when only coupled to a single bath. 
Using that the coupling between the spins and the baths is weak, it is possible to compute the energy exchanged with the baths during one period of evolution, $T_{\rm d}$, by $Q=\int_{t_0}^{t_0+T_{\rm d}}\tr\!\!\spare{\Hopi\Dop\pare{\rhop}}dt$, where $t_0$ is a sufficiently long time to reach asymptotic periodic state. Since the dynamics of the entire system is periodic, this holds in particular for the energy of the spins. This implies that $\langle \sum_j\frac{\hbar\w_j}{2} \sop_{z,j} \rangle$ also has period $T_{\rm d}$ and the net energy absorbed by the spins over one period vanishes. Hence, all energy absorbed from the baths is transferred to the load $Q=-W$, where 
\begin{align}
W=\sum_j \int_{t_0}^{t_0+T_{\rm d}}\hbar g_j\;\tr\!\!\!\spare{   \frac{d}{dt} \left(\ave{\dop\,}\sopp_j + \ave{\ddop} \sopm_j\right) \rhop_{s,j}} dt
\end{align}
is the energy exchanged with the effective periodic driving $\ave{\dop\,}$. Using Eq.~(\ref{eq:dxopdt}), one can show that $W$ is naturally related to the energy increase of the particle by 
\begin{align}
W=T_{\rm d}\hbar\nu\left.\frac{d\ave{\xop}}{dt}\right|_{\rm sep}. \label{eq:W}     
\end{align}  
Eq.~(\ref{eq:W}) states that all the energy absorbed by the bath is used to drive the particle on the lattice. However, this analysis of the set-up as a continuously driven engine described by the separable ansatz in Eq.~(\ref{eq:sep}) does not accurately describe the energy and entropy exchanges between the system and the load in all parameter regimes. A more accurate description is obtained by including the load (and its correlations with the engine) into the analysis.

{\it Conclusions}: In microscopic engines it is possible that some non-negligible entropy transfer occurs between the engine and the load. This can be studied accurately for self-contained engines, in which the full quantum state of the engine and the load are considered exactly without any separable assumption. This entropy transfer has important consequences, in fact, as similarly discussed in \cite{Scully2001}; it allows the engine to push the particle up the lattice, despite the presence of only a single bath.  

Self-contained engines have another advantage concerning the characterization of their performance. While it is not possible to use the definition of work as carefully described in \cite{TalknerHanggi2007, CampisiTalkner2011,CampisiTalkner2011err} because the system is time independent and always in contact with a bath, it is however possible to evaluate how far and how quickly the particle moves up the tilted lattice. This was already used by Carnot to describe the motive power of an engine. In fact in \cite{Carnot1824} he wrote: ``We use here the expression motive power to express the useful effect that a motor is capable of producing. This effect can always be likened to the elevation of a weight to a certain height.'' 

For our self-contained system, since after a time of a few tens of $2\pi/\nu$, the motion of the particle is well described by classical diffusive dynamics, repeated measurements at long enough intervals will result in an average linear displacement of the particle at the velocity found in our work. More frequent measurements would instead result in a quantum Zeno-like behaviour for which the velocity of the particle would be highly suppressed.

{\it Acknowledgments}: We are thankful to J.M. Arrazola, G. Benenti, A. Mari and A. Roulet for fruitful disucssions. D.P. acknowledges fundings from Singapore MOE Academic Research Fund Tier-2 project (Project No. MOE2014-T2-2-119, with WBS No. R-144-000-350-112), together with U.B., from SUTD-MIT IDC (Project No. IDG21500104), together with C.T. and from AcRF MOE Tier-I (project SUTDT12015005).

\begin{appendix}

\section{Solution of spin dynamics for separable ansatz}
From the spin equation of motion, Eq.~\eqref{eq:ucs}, we can write a set of coupled equations of motion for $\ave{\sigma_j^- (t)},\ave{\sigma_j^+(t)}$ and $\ave{\sigma_{z,j}}$. The resulting time-dependent equations can be easily solved by transforming into a rotating frame via the transformation, $\ave{\tilde \sigma_j^-(t)} = \ave{\sigma_j^-(t)} e^{\im (\nu t+\phi )}$, where $\phi$ is given by $d_0=|d_0|e^{-\im\phi}$. This results in the matrix equation,
\begin{align}
\partial_t \begin{pmatrix}
\ave{\tilde \sigma^-_j}\\
\ave{\tilde \sigma^+_j}\\
\ave{\sigma_{z,j}}
\end{pmatrix}
= \hat M_j \begin{pmatrix}
\ave{\tilde \sigma^-_j}\\
\ave{\tilde \sigma^+_j}\\
\ave{\sigma_{z,j}}
\end{pmatrix} +
\begin{pmatrix}
0\\
0\\
\lambda^+ - \lambda^-
\end{pmatrix},
\end{align}
where the time \emph{independent} matrix $\hat M_j $ is given by 
\begin{align}
\hat M_j &= \begin{pmatrix}
-\half{\lambda_j} + \im \pare{\w_j -\nu}& 0 & -\im g |d_0| \\
0 & -\half{\lambda_j}- \im \pare{\w_j -\nu} & \im g |d_0| \\
-2\im g |d_0| & 2\im g |d_0|  & - \lambda_j \\
\end{pmatrix}.
\end{align}
The steady state solution to this equation can then be found by setting the time derivatives to zero and solving the resulting system of linear equations. In particular, the solution for $\ave{\tilde \sigma_j^+}$ in the steady state is
\begin{align}
\lim_{t\to\infty}\ave{\tilde \sigma_j^+} =\pare{\frac{\lambda_j^+ - \lambda_j^-}{\lambda_j}} \frac{2|d_0|g\pare{2\pare{\w_j - \nu} - \im \lambda_j}}{8|d_0|^2 g^2 + \lambda_j^2 + 4\pare{\w_j - \nu}^2}.
\end{align}

\section{Derivation of the diffusion equation}
Let us consider a density operator in Eq.~(\ref{eq:null}). Applying the perturbation $\Vp$ gives 
\begin{align}
\spare{\Vp\Pro_0}(\rhop)=-\im g\sum_{n,j} & P_n \left\{\kb{n+1}{n} \otimes p^+_j \kbj{\downarrow}{\uparrow}  \right.  \nonumber \\
& + \kb{n-1}{n} \otimes_j p^-_j \kbj{\uparrow}{\downarrow}   \nonumber \\ 
& - \kb{n}{n+1} \otimes_j p^+_j \kbj{\uparrow}{\downarrow}   \nonumber \\ 
& - \left.\kb{n}{n-1} \otimes_j p^-_j \kbj{\downarrow}{\uparrow} \right\} \otimes \Ss_j.
\end{align}
Here we use the notation that all the superoperators between square brackets act, from right to left, on the density operator in the round bracket to their right. We have also used the notation $\Ss_j=\otimes_{i\ne j} \pare{p^+_j\proj{\uparrow}+p^-_j\proj{\downarrow}}$.        
Subsequently, $(-\Lz)^{-1}$ results in 
\begin{align}
&\spare{(-\Lz)^{-1}\Vp\Pro_0}(\rhop)=\im g\sum_{n,j}  P_n \nonumber \\ 
&\times\left\{-\frac{1}{\im(\Del-\w_j)+\lt_j}\kb{n+1}{n} \otimes_j p^+_j \kbj{\downarrow}{\uparrow} \right.  \nonumber \\
& + \frac{1}{\im(\Del-\w_j)-\lt_j}\kb{n-1}{n} \otimes_j p^-_j \kbj{\uparrow}{\downarrow} \nonumber \\ 
& - \frac{1}{\im(\Del-\w_j)-\lt_j}\kb{n}{n+1} \otimes_j p^+_j \kbj{\uparrow}{\downarrow} \nonumber \\ 
& + \frac{1}{\im(\Del-\w_j)+\lt_j}\left.\kb{n}{n-1} \otimes_j p^-_j \kbj{\downarrow}{\uparrow} \right\} \otimes \Ss_j           
\end{align}
and finally, applying the perturbation $\Vp$ a second time followed by a projection on the null space of $\Lz$, gives 
\begin{align}
\spare{\Pro_0\Vp(-\Lz)^{-1}\Vp\Pro_0}&(\rhop)=\nonumber \\ 
\sum_{n,j} \frac{ 2 g^2\lt_j P_n}{(\Del-\w_j)^2} &  \left\{ \proj{n+1} \otimes_j p^+_j \downprojj\right.  \nonumber \\
& - \kb{n}{n} \otimes_j \pare{p^+_j \upprojj+p^-_j \downprojj} \nonumber \\ 
& + \left.\proj{n-1} \otimes_j p^-_j \upprojj \right\}  \otimes \Ss_j.  
\end{align}
After tracing out the spins degrees of freedom we obtain Eq.~(\ref{eq:padi}).   
\end{appendix}

%%%%%%%%%%%%%%%%%%%%%%%%%%%%%%%%%%%%%%%%%%%%
%%%%%%%%%%%%%%%%%%%%%%%%%%%%%%%%%%%%%%%%%%%%
%bibliography    


\begin{thebibliography}{10}
\bibitem{lowerbound} We have also studied semi-infinite lattices bounded from below and found asymptotic similar dynamics as the one discussed in the following.
\bibitem{initialcondition} In most cases we set, at $t=0$, the spins in the $|\! \downarrow\ran_j$ state, while the particle is in a pure state with a Gaussian position distribution. We have however also considered more complex and correlated initial conditions without noticing any qualitative difference in the asymptotic dynamics.
\bibitem{validity} The adiabatic elimination is valid as long as $g_j\ll \lt_j$.
\bibitem{shorttime} At short times peaks at harmonic frequencies are present, but as time progresses the peaks disappear leaving only the central one.
\end{thebibliography}
\end{document}